\documentclass[rnote]{aa} 
\usepackage{graphicx}
\usepackage{txfonts}
\usepackage{natbib}
\begin{document}
\title{Searching for X-ray emission from AGB stars}
\author{S. Ramstedt\inst{1} \and R. Montez\inst{2} \and J. Kastner\inst{2} \and W.~H.~T. Vlemmings\inst{3}}
\offprints{S. Ramstedt}
\institute{Argelander Institute for Astronomy, University of Bonn, 53121 Bonn, Germany \\ \email{sofia@astro.uni-bonn.de} \and Rochester Institute of Technology, 54 Lomb Memorial Drive, Rochester, NY 14623, USA \and Onsala Space Observatory, 439 92 Onsala, Sweden}    
\date{Received; accepted}
\abstract{Magnetic fields have been measured around Asymptotic Giant Branch (AGB) stars of all chemical types using maser polarization observations. If present, a large-scale magnetic field would lead to X-ray emission, which should be observable using current X-ray observatories.}{The aim is to search the archival data for AGB stars that are intrinsic X-ray emitters.}{We have searched the ROSAT, CXO, and XMM archives for serendipitous X-ray observations of a sample of $\sim$\,500 AGB stars. We specifically searched for the AGB stars detected with GALEX. The data is calibrated, analyzed and the X-ray luminosities and temperatures are estimated as functions of the circumstellar absorption. }{We identify 13 AGB stars as having either serendipitous or targeted observations in the X-ray data archives, however for a majority of the sources the detailed analysis show that the detections are questionable. Two new sources are detected by ROSAT: T~Dra and R~UMa. The spectral analysis suggests that the emission associated with these sources could be due to coronal activity or interaction across a binary system.}{Further observations of the detected sources are necessary to clearly determine the origin of the X-ray emission. Moreover, additional objects should be subject to targeted X-ray observations in order to achieve better constraints for the magnetic fields around AGB stars.}  
\keywords{Stars: AGB and post-AGB  -- Stars: magnetic field -- X-rays: stars}
   
\maketitle

  
\section{Introduction}
\label{intro}
The required conditions for AGB stars to evolve into planetary nebulae is still a matter of debate. Since AGB circumstellar envelopes (CSEs) are normally spherically symmetric \citep{castetal10} and PNe are not \citep{parketal06}, the shaping during the transition is one of the open questions. One suggested shaping agent is a large-scale, circumstellar magnetic field. Circular polarization due to magnetic-field-induced Zeeman splitting has been measured using maser polarization observations \citep[e.g.,][]{herpetal06,vlemetal05,bainetal03}. For C-type (C/O\,$>$\,1) stars, Zeeman splitting of the paramagnetic molecule CN has been used to estimate the magnetic field \citep{herpetal09}, and recently, the first estimate of a magnetic field around an evolved star using polarized CO emission was published \citep{vlemetal12}. The current measurements are consistent with either an $R^{-1}$ (toroidal) or an $R^{-2}$ (poloidal) dependence, indicating field strengths of order several tens of Gauss at the stellar surface \citep{vlem10}.

A large-scale magnetic field will likely lead to X-ray emission \citep{blacetal01,pevtetal03}. However, a large fraction, if not all, of this emission might be absorbed by the high-density wind surrounding the star \citep{kastsoke04a}. Targeted X-ray observations have to our knowledge only been performed for three AGB stars: $o$ Cet (Mira), T~Cas, and TX~Cam. Known symbiotic systems with AGB primaries, e.g., R~Aqr and CH~Cyg, have also been observed \citep{kelletal01,karoetal07}. \citet{kastsoke04b} observed the Mira AB system with XMM and conclude that the weak, predominantely soft ($<3$\,keV) X-ray emission probably originates from the accretion from Mira A to B, or from coronal activity on Mira B. Later \citet{karoetal05} reported emission at both binary components, and the detection of a large X-ray outburst on Mira A, with CXO. The spectrum was dominated by a soft ($0.2-0.7$\,keV) component originating from Mira A. The total luminosity was roughly a factor of 2 to 4 higher than that of the previous observations. T~Cas and TX~Cam were both observed with XMM, but neither was detected \citep{kastsoke04a}, possibly due to the absorption of the high density wind.

We have searched the ROSAT, CXO, and XMM archives for X-ray sources among a sample of $\sim$\,500 well-studied, nearby AGB stars and will report our findings here. In Sect.~\ref{data} the sample is presented and the observations are described. The results are given in Sect.~\ref{res} and discussed in Sect.~\ref{dis}. Finally, we draw our conclusions in Sect.~\ref{con}. 


\section{Data and analysis} 
\label{data}

\subsection{The sample}
\label{samp}
We have searched the archives for X-ray detections associated with the 180 AGB stars in the samples of \citet{schoolof01}, \citet{delgetal03}, and \citet{ramsetal06}. The C-type star sample is complete, or close to complete, out to 500\,pc. The S-type (C/O\,$\sim$\,1) star sample is a close to complete sample of mass-losing S-type stars out to 600\,pc. The completeness of the M-type (C/O\,$<$\,1) sample has not been thoroughly investigated. To supplement the samples, we added nine stars found to have far-UV excess with GALEX \citep{sahaetal08} and 291 M-type Miras of \citet{littmare90}. Thirteen sample stars were found to have either serendipitous or targeted observations in the archives. These stars are discussed in detail in the paper, and are presented in Table~\ref{sample} with basic stellar parameters.

\begin{table}[htdp]
\caption{The AGB stars discussed in detail in the paper. References for the distance estimates are given in the footnote.}
\begin{center}
\begin{tabular}{lllccc}
\hline \hline
Source \hspace{2cm} & Sp.  & Var. &  $P^{1}$ & $L^{2}$ & $D^{3}$  \\ 
	& type & type & [d] & [$L_{\odot}$] & [pc] \\
\hline
\object{UX Ara} & $\cdots$ & M & 250 & \phantom{1}4570 & 1000: \\
\object{T Cas} & M & M & 445 & \phantom{1}9200 & \phantom{1}290$^{4}$ \\
\object{TX Cam} & M & M & 557 & 11900 & \phantom{1}440$^{5}$ \\
\object{$o$ Cet} & M & M & 332 & \phantom{1}6430 & \phantom{12}92$^{6}$\\
\object{W Cyg} & M & SRb & 131 & $\cdots$ & \phantom{1}173$^{6}$ \\
\object{R Dor} & M & SRb & 338 & $\cdots$ & \phantom{12}55$^{6}$ \\
\object{T Dra} & C & M & 422 & \phantom{1}6550 & \phantom{1}610$^{7}$ \\
\object{RT Eri} & M & M & 371 & \phantom{1}7380 & \phantom{1}277$^{8}$ \\
\object{R Leo} & M & M & 310 & \phantom{1}5920 & \phantom{12}71$^{6}$\\
\object{L$_{2}$ Pup} & M & SRb & 141 & $\cdots$ &  \phantom{12}64$^{6}$ \\
\object{RW Sco} & M & M & 388 & \phantom{1}7800 & \phantom{1}870: \\
\object{R UMa} & M & M & 302 & \phantom{1}5750 & \phantom{1}415$^{6}$ \\
\object{SS Vir} & C & SRa & 364 & $\cdots$ & \phantom{1}540$^{7}$ \\ 
\hline
\end{tabular}
\end{center}
{\tiny $^{1}$ \citet{samuetal09}; $^{2}$ \citet{whitetal94,groewhit96}; 
$^{3}$ When a reliable parallax ($p/\Delta p$\,$>$\,2) is available, it has been used to estimate the distance. For the remaining objects we have estimated the distance using the K-band magnitude \citep{cutretal03} and PL relation \citep{whitetal08} (marked by a colon); 
$^{4}$ \citet{loupetal93}; $^{5}$ \citet{ramsetal08}; $^{6}$ \citet{vanL07}; $^{7}$ \citet{schoolof01}; $^{8}$ \citet{knapetal03} }
\label{sample}
\end{table}%


\subsection{ROSAT observations}

\subsubsection{Data description and preparation}

Among the AGB star sample listed in Table~\ref{sample}, only 3 objects --- T~Dra, R~UMa, and $o$~Cet (Mira)--- have spatially coincident X-ray sources listed in the ROSAT source catalogs maintained at HEASARC\footnote{http://heasarc.nasa.gov/} (Table~\ref{ptobs}). X-ray sources very near the positions of T~Dra and R~UMa were detected by the ROSAT position sensitive proportional counter (PSPC, in scanning mode) during the ROSAT All-Sky Survey (RASS). These sources are $\sim$\,8\arcsec \,and $\sim$\,2\arcsec \, from the catalogued positions of T~Dra and R~UMa, respectively, and hence within the 10\arcsec \, positional error. The pointed ROSAT archival observation of Mira \citep{sokekast03} is also included in the following analysis.  

\begin{table}[htdp]
\caption{Observations of the sample objects.}
\begin{center}
\begin{tabular}{lcccccc}
\hline \hline
Object & Obs.$^{1}$ & ObsID &  Obs. Date & $t_{\textrm{exp}}$ (ks) &   $\theta$ ($^{\prime}$)$^{2}$  \\ 
\hline
UX~Ara & XMM-S & 0306171201 & 2006-03-01 & \phantom{1}6.1 & 13.2 \\ 
T~Cas & XMM-T& 0148500501 & 2003-02-06 & 13.2 & \phantom{1}1.1 \\
TX~Cam & XMM-T & 0148500101 & 2003-09-04 & 13.9 & \phantom{1}1.1 \\
$o$~Cet & ROSAT & 201501 & 1993-07-15 & \phantom{1}9.1 & \phantom{1}0.3\\
W~Cyg & XMMSS & 9146000003 & 2007-11-29 & $\sim$\,10$^{-2}$ & \phantom{1}0.1\\
R~Dor &  XMMSS & 9115800003 & 2006-04-06 & $\sim$\,10$^{-2}$ & \phantom{1}0.2\\
T~Dra & RASS & 930730 & 1990-07-30 & \phantom{1}3.0 & \phantom{1}0.1 \\
RT~Eri & CXO-S & 4064 & 2003-06-17 & \phantom{1}4.4 & \phantom{1}1.8 \\
R~Leo &  XMMSS & 9126500005 & 2006-11-06 & $\sim$\,10$^{-2}$ & $<$\,0.1\\
L$_{2}$~Pup &  XMMSS & 9077600003 & 2004-03-05 & $\sim$\,10$^{-2}$ & $<$\,0.1\\
RW~Sco & XMM-S & 0554440101 & 2008-08-26 & 30.1 & \phantom{1}9.1 \\ 
R~UMa & RASS & 930513 & 1990-10-05 & \phantom{1}0.5 & $<$\,0.1 \\
SS~Vir & XMM-S & 0303560101 & 2005-07-10 & \phantom{1}4.4 & 13.8\\ 
\hline
\end{tabular}
\end{center}
{\tiny $^{1}$ The XMM slew survey is denoted XMMSS. Whether the observations were serendipitous (S) or targeted (T) is also indicated. \\
$^{2}$ $\theta$ is the off-axis angle of the object with respect to the aimpoint of the observation. }
\label{ptobs}
\end{table}

New calibration and analysis (Sect.~\ref{ss:spec}) were applied to the ROSAT archival observations. The newly determined background-subtracted count rates are 20.6\,counts\,ks$^{-1}$ for T~Dra, 39.7\,counts\,ks$^{-1}$ for R~UMa, and 6.7\,counts\,ks$^{-1}$ for Mira. According to the RASS faint source catalogue (FSC), the extent of the X-ray emission from T~Dra is significantly larger than the PSPC point spread function (PSF), which suggests extended emission or a superposition of 2 or more X-ray point sources. 

\begin{figure}[]
\begin{center}
\includegraphics[width=\columnwidth]{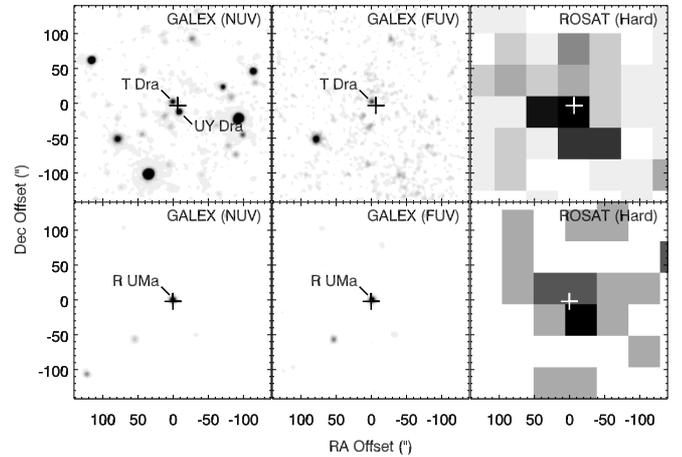}
\caption{The GALEX NUV (1771--2831\,\AA, left) and FUV (1344--1786\,\AA, middle), together with the ROSAT X-ray images (right) of T~Dra (upper panel) and R~UMa (lower panel). The crosses correspond to the coordinates of the X-ray sources according to the RASS.}
\label{pos}
\end{center}
\end{figure}

To further examine the fields containing T~Dra and R~UMa, the available GALEX observations \citep{sahaetal08} were obtained from the GALEX archive. The ROSAT X-ray images are presented in Fig.~\ref{pos} together with the UV images\footnote{The spatial resolution of GALEX is 4.3\arcsec and 5.3\arcsec in the FUV and NUV, respectively, \citep{morretal07}}. UY~Dra \citep[spectral type K2III--IV, ][]{samuetal09}, also seen in the NUV image in Fig.~\ref{pos} (upper panel, right), is located less than 20\arcsec south-west of T~Dra. Although UY~Dra is the less likely X-ray source (since it's not visible in the FUV image), new X-ray observations at higher sensitivity and spatial resolution are required to draw firm conclusions as to whether T~Dra or UY~Dra, or possibly both, is emitting X-rays. 

\subsubsection{Spectral analysis} 
\label{ss:spec}
Source and background spectra were extracted from the ROSAT observations.  
The exposure times for the spectra are adjusted according to the observation-specific exposure maps. New ancillary response files were created with the HEASoft tool {\it pcarf} using the canned response matrix file appropriate for these PSPC observations. Bad channels were removed, and the remaining channels are grouped to increase the signal to noise. Fitting of the background-subtracted spectra was performed in XSPEC (version 12.6.0q). Churazov weighting of the spectral bins was used to improve the fit statistics.  An absorbed optically-thin thermal plasma ({\it wabs*(raymond)}) was assumed as the spectral model. The column density, $N_{\rm{H}}$, is used to estimate the absorption due to H, H$_{2}$ (assuming equal amounts of atomic and molecular hydrogen), and other atomic species, assuming an optically thin thermal plasma with solar abundances. The low count spectra and ROSAT sensitivity do not allow for the determination of the absorbing column density; instead a range of values of $N_{\rm{H}}$ was assumed while the plasma temperature and luminosity were allowed to vary during the fitting procedure. 

\subsection{Pointed XMM and CXO observations}
In addition to XMM and CXO observations targeting T~Cas, TX~Cam, and Mira \citep{kastsoke04a, kastsoke04b, karoetal05}, four sample objects have been serendipitously observed in pointed observations by these two observatories. UX~Ara, RT~Eri, and RW~Sco are not detected in these observations. SS Vir is 13.3$^{\prime}$ off-axis in its serendipitous XMM observation and falls on the buffer region removed during pipeline processing. In the unfiltered observation, there is apparently a tentative detection at the position of SS Vir. In the case of the pointed XMM observation of T~Cas, determined to be an X-ray non-detection by \citet{kastsoke04a}, a closer inspection of the data showed that a potential X-ray source is marginally detected in one out of the three XMM detectors. However, we conclude that the apparent detections of SS~Vir and T~Cas are likely due to optical loading and are hence both spurious (see Appendix~\ref{opticalloading}).

\subsection{XMM slew observations} 
Four sources in the XMM slew survey appear to be associated with AGB stars: R~Dor, W~Cyg, R~Leo, and L$_{2}$~Pup. They are free of quality warning flags and appear only in the soft (0.2-2\,keV) energy band. The raw data was reprocessed and the examination of the spectra and event pattern distributions, together with the visual brightness of the sources, lead us to conclude that the detections are false, and due to optical loading (Appendix~\ref{opticalloading}).


\section{Results}
\label{res}

\subsection{X-ray spectral distributions and luminosities}
\label{ss:fltem}

\begin{figure}[]
\begin{center}
\includegraphics[width=\columnwidth]{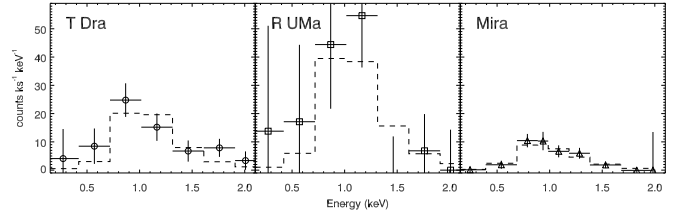}
\caption{The ROSAT X-ray spectra of T~Dra, R~UMa, and Mira together with typical model fit (dashed). Mira did not appear in the RASS BSC or FSC, since its count rate is lower than the two FSC objects T~Dra and R~UMa.}
\label{spectra}
\end{center}
\end{figure}

The spectra of T~Dra and R~UMa are shown in Fig.~\ref{spectra} together with the reanalyzed ROSAT spectrum of Mira and a model assuming $N_{\rm H}$\,=\,$5\times10^{21}$\,cm$^{-2}$ and $T_{\rm X}$\,$\sim$\,10$^{7}$\,K. The spectra are not very sensitive to the assumed values and the overlaid models should therefore be considered only as representative. The error bars are weighted Poisson errors. The observed emission peaks at around 1\,keV in all three sources. 

\begin{figure}[]
\begin{center}
\includegraphics[width=\columnwidth]{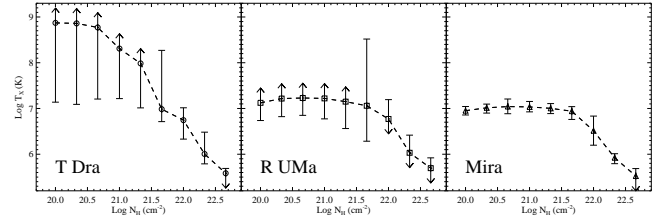}
\caption{The best-fit X-ray temperatures as functions of the hydrogen column density for T~Dra (squares), R~UMa (dots), and Mira (triangles). The 90\% confidence range is indicated by the flat-tipped error bars, while an unconstrained value is indicated by the arrow-tipped error bars. The behavior of the fits reflects the limitations of ROSAT spectra for hard and low count rate sources.}
\label{fluxtx}
\end{center}
\end{figure}

The fits to the ROSAT spectra demonstrate the difficulty of simultaneously constraining the temperature and column density for the AGB sources.  However, we can use the model predictions to determine the reliability of the fitted parameters. For all values of column density, almost all of the best-fit models produce acceptable fit statistics (reduced $\chi^2$ between 0.8 and 1.6), but the temperature is often unconstrained at the 90\% confidence level. Figure~\ref{fluxtx} displays the best-fit temperatures for each AGB source versus the assumed absorbing column density. For low column densities, the temperature becomes unconstrained, but the lower 90\% confidence range is consistently around 10$^{7}$\,K, as expected for coronal or compact object accretion processes. 

\begin{figure}[]
\begin{center}
\includegraphics[width=\columnwidth]{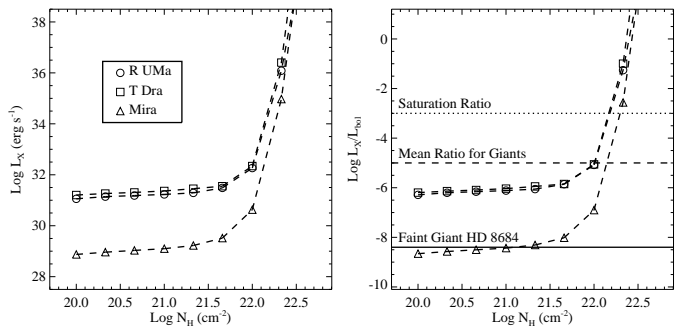}
\caption{The X-ray luminosities, L$_{\rm{X}}$ (left), and L$_{\rm{X}}$/L$_{\rm{bol}}$ (right) as functions of $N_{\rm{H}}$ for T~Dra, R~UMa, and Mira. In the right plot, the horizontal lines mark the typical values for coronal saturation around young and active stars \citep[$10^{-3}$;][]{feigetal03}, the mean activity ratio for G-K giants \citep[$10^{-5}$;][]{gond05,gond07} and the lowest level that has been detected in a first-ascent giant \citep{huenetal96}. }
\label{lumx}
\end{center}
\end{figure}

The unabsorbed source flux is predicted from the model fits with the absorbing column removed, and used along with the distances in Table~\ref{sample} to determine the X-ray luminosity in the 0.2 to 2.0\,keV energy band as a function of the assumed column density (Fig~\ref{lumx}). The X-ray luminosity scaled to the bolometric luminosity of the AGB star provides a strong diagnostic of the physical process responsible for the X-ray emission. These results and their constraints on the column density and origin of the X-ray emission are discussed further in Sect.~\ref{dis}. 

\section{Discussion}
\label{dis}

\subsection{Estimating the absorbing column density: $N_{\rm{H}}$}

Based on the luminosity levels shown in Fig.~\ref{lumx}, a column density much higher than $10^{22}$\,cm$^{-2}$ seems unlikely for the sources detected by ROSAT. By following \citet[][their Eqn.~1]{kastsoke04a}, the hydrogen column density of the circumstellar wind, $N_{\rm{H}}$, can be estimated from the mass-loss rate, $\dot{\rm{M}}$, and the expansion velocity, $v_{\rm{e}}$. Using the parameters from \citet{schoolof01}, the circumstellar hydrogen column density of T~Dra is 8.9$\times10^{22}$\,cm$^{-2}$. Adopting typical values for M-type stars ($\dot{M}$\,=\,10$^{-7}$\,$M_{\odot}$\,yr$^{-1}$, $v_{\rm{e}}$\,=\,7\,km\,s$^{-1}$) yields $N_{\rm{H}}$\,$\approx$\,10$^{22}$\,cm$^{-2}$, as an approximate estimate for R~UMa. We have compared the results of the equation given by \citet{kastsoke04a} to estimates from measurements of the 21cm-emission of atomic hydrogen around AGB stars \citep{geralebe06,mattetal11} by calculating the hydrogen column densities from the estimated mass and extent of the emission, assuming an $r^{-2}$-density distribution. The results (Appendix~\ref{nh}) show no indication that the equation systematically overestimates the hydrogen column density as suggested by \citet{kastsoke04a}. 

If the AGB star is the source of X-ray emission, the detections might indicate that Eqn.~1 slightly overestimates $N_{\rm{H}}$, but more likely it reflects the simplicity of the analysis. It is to be expected that a process that could be responsible for the X-rays (i.e.~binary interaction or a strong magnetic field) would also introduce asymmetries in the circumstellar environment. The emission could therefore emerge from a region where the column density of the AGB envelope falls below a few times 10$^{22}$\,cm$^{-2}$, although the average column density as probed by unresolved observations, is higher. These uncertainties can only be mitigated by additional observations (see Sect.~\ref{abs_other}).

\subsection{X-ray emission from a potential companion}
\label{comp}

The suspicion of undetected binary companions to R~UMa and T~Dra stems from the multiplicity flag (`X') in their Hipparcos catalog entries. This flag indicates that neither a single or binary astrometric solution could be found and may suggest that the star is a binary \citep{lindetal97}.  Accretion of the AGB wind onto a compact companion can account for the X-ray emission, e.g., the emission detected from Mira B \citep{kastsoke04b} is believed to be due to accretion onto a compact white dwarf (WD). Furthermore, accretion onto a late-type companion can rejuvenate its corona leading to hard X-ray emission \citep[e.g.][]{montezetal10,jefste96}.

\citet{sahaetal08} searched for companions to AGB stars with GALEX UV observations in a sample of stars largely selected on the Hipparcos multiplicity flag. R~UMa and T~Dra were both found to have UV excesses. Only R~UMa was modeled since spectral templates did not exist for carbon-rich AGB stars, e.g. T~Dra. \citet{sahaetal08} find that the UV excess of R UMa may be explained by a hot companion at $\sim$\,9000\,K. Their model provides the luminosity relative to the primary star, and assuming a distance to R~UMa of 0.5\,kpc results in a companion luminosity of L$_{c}$\,=\,0.85\,$L_{\odot}$. This luminosity does not agree with evolutionary tracks for a cooling WD and the origin of the UV excess therefore remains uncertain. The properties determined by \citet{sahaetal08} suggests it is unlikely that accretion onto a WD is responsible for the X-ray emission. Also, in light of these potential X-ray detections, it is possible that the UV excess is unrelated to the photosphere of an unknown companion, but instead due to chromospheric activity around a late-type companion or the AGB star itself.  Such an interpretation is consistent with the X-ray emission being caused by coronal emission. 

Indeed, the hard X-ray emission detected from WDs is typically attributed to the coronal emission of their late-type main sequence companions \citep{odwyetal03,chuetal04,bilietal10}. A coarse measure of the spectral distribution is provided by the hardness ratios (HR). The ROSAT PSPC HRs are defined as described in, e.g., \citet{kastetal03}. HR1 is related to the soft part of the spectrum, and a small HR1 indicates a very soft spectrum. HR2 is related to the harder part of the spectrum, and if HR2 is large, the spectrum is hard. The binary WDs detected by ROSAT have consistent HR2 ratios, but only a few reach the high HR1 ratios found from the AGB sources. The HR2 ratios are consistent with high temperature ($>$\,10$^{6}$\,K) plasma that may be due to coronal or accretion processes. The high HR1 ratios indicate a large absorbing column which attenuates the soft X-ray emission, as expected from the dense AGB mass loss.  If we assume that the coronae of unknown companions in T~Dra and R~UMa are saturated and responsible for the X-ray emission ($L_{X}/L_{\rm bol}$\,$\sim$\,10$^{-3}$), then their luminosities should be $\sim$\,10\,$L_{\sun}$ (if $N_{\rm{H}}$\,=\,1$\times$10$^{21}$\,cm$^{-2}$). Whereas, if the AGB star is responsible for the X-ray emission (see next section), the $L_{X}/L_{\rm bol}$ ratios in Fig~\ref{lumx} are similar to those observed in late-type giants \citep[$10^{-5}$;][]{gond05,gond07}. 

UY~Dra could possibly be the source of the X-ray emission in the vicinity of T~Dra, but is not visible in the FUV GALEX image whereas T~Dra is clearly present. Furthermore, our analysis shows that the extent of the X-ray emission is larger than the ROSAT PSF, perhaps indicating multiple X-ray sources. 

\subsection{X-ray emission from the AGB star}

The pulsation-induced shocks above the AGB atmosphere will not emit X-rays. Typical post-shock temperatures are $\lesssim$\,15000\,K leading to radiative cooling mainly through forbidden lines and hydrogen lines in the optical and UV \citep{schietal03}. Thus, if a single AGB star is the source of X-ray emission, one possible explanation is coronal emission from a large-scale magnetic field. The origin and feasibility of a large-scale magnetic field on an AGB star is discussed in several papers \citep[e.g.,][]{garcetal05,sokekast03} and could be due to a dynamo \citep{blacetal01}, the movement of ionized gas in large convective cells \citep{huesch96}, or the presence of a binary companion as in the cool components of symbiotic systems \citep{soke02b,zamaetal08}. 

\citet{pevtetal03} found a correlation between magnetic flux and $L_X$ using observations of the Sun and magnetically active stars. Assuming $N_{\rm{H}}$\,=1$\times$10$^{21}$\,(10$^{22}$)\,cm$^{-2}$, the magnetic flux would be $\approx$\,10$^{23}$\,(10$^{24}$)\,G\,m$^{2}$. If we assume a stellar radius of 1\,AU and that the X-rays are emitted evenly across the surface, this gives a surface magnetic field of $\sim$\,1\,G ($\sim$\,10\,G), which agrees with typical values extrapolated from maser polarimetry measurements of other AGB stars \citep[e.g.,][]{vlem10}. 

\subsection{Questions to be addressed by future observations}
\label{abs_other}

The uncertainties related to the absorption of the X-ray emission are many and due to uncertainties in the composition, the atomic/molecular fraction, the dust-to-gas ratio, as well as the effects of circumstellar asymmetries. Spatially resolved observations of the gas and dust distribution around the X-ray detected sources would be needed to evaluate if the detections are the result of less absorption along the line of sight to these sources.

\begin{figure}[]
\begin{center}
\includegraphics[width=\columnwidth]{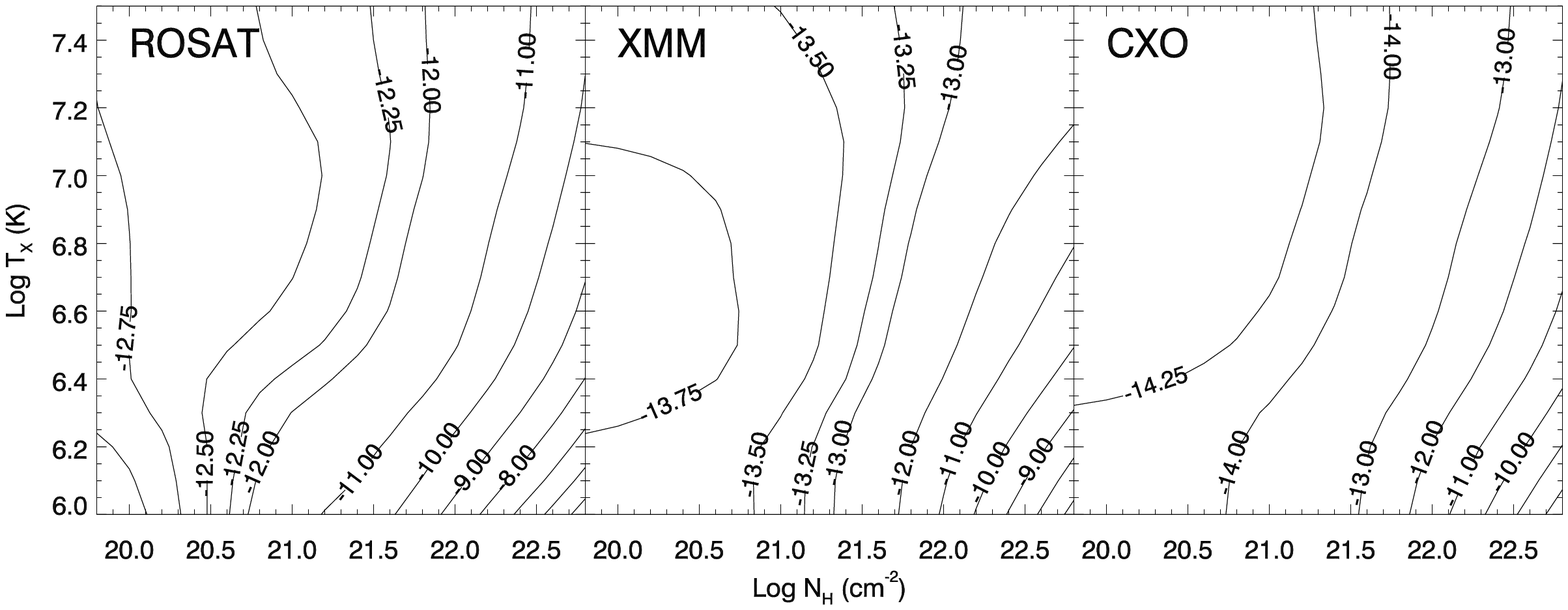}
\caption{The detection thresholds (log($F_{X}$) in ergs\,cm$^{-2}$\,s$^{-1}$) for combinations of plasma temperatures and $N_{\rm{H}}$ for ROSAT, XMM, and CXO.}
\label{agbfuture}
\end{center}
\end{figure}

Further X-ray observations are necessary to place meaningful limits on magnetic fields around AGB stars. In Fig.~\ref{agbfuture} we show the detection thresholds (contours at log($F_{X}$)) for combinations of plasma temperatures and $N_{\rm{H}}$ expected from coronal or accretion-related emission from AGB stars or their companions for ROSAT, XMM, and CXO. The detection threshold is determined from the count rates of 20\,counts\,ks$^{-1}$ for ROSAT, 10\,counts\,ks$^{-1}$ for XMM (thick filter), and 1 count ks$^{-1}$ for CXO. The flux levels of T~Dra and R~UMa are $F_{X}$\,$\approx$\,10$^{-12}$\,ergs\,cm$^{-2}$\,s$^{-1}$, and from Fig.~\ref{agbfuture} it is clear that with CXO we can detect emission at this level even if it is heavily absorbed, i.e. $N_{\rm{H}}$\,$>$\,10$^{22}$\,cm$^{-2}$.

\section{Conclusions}
\label{con}
We have searched the ROSAT, XMM, and CXO archives for X-ray detections of a large sample of Galactic AGB stars. Thirteen observations of AGB stars are found. The targeted observations of TX~Cam, T~Cas, and Mira have previously been analyzed and discussed in \citet[][TX~Cam and T~Cas]{kastsoke04a}, and \citet[][Mira]{karoetal05}. Four AGB stars, W~Cyg, R~Dor, R~Leo, and L$_{2}$~Pup were apparently detected as part of the XMMSS, however, the analysis of the data shows that the detected counts are likely due to optical photons (Appendix~\ref{opticalloading}). We urge caution when considering visually bright and very red sources discovered in the XMM Slew Survey catalogs. Four additional AGB stars, UX~Ara, RT~Eri, RW~Sco, and SS~Vir, were observed serendipitously by XMM or CXO. Three were not detected, and a closer inspection of the data on SS~Vir showed that the detection is due to optical loading (Appendix~\ref{opticalloading}).

We finally conclude that two of the sample AGB stars, the carbon star T~Dra and the M-type star R~UMa, likely emit X-rays. The spectra of both sources peak at $\approx$1\,keV, i.e. at the same energy as the ROSAT spectrum of Mira AB. The low count rate and ROSAT sensitivity does not allow for a secure determination of the absorption or the X-ray luminosity, but we find that the emission is consistent either with accretion onto a WD companion or with coronal emission from a magnetic field of the same order of magnitude as has been inferred from maser observations of AGB stars. To better evaluate the origin of the emission, targeted XMM or CXO observations will be necessary. 


\begin{acknowledgements}
The authors would like to thank the anonymous referee for helpful suggestions that greatly improved the manuscript. SR acknowledges support by the Deutsche Forschungsgemeinschaft (DFG) through the Emmy Noether Research grant VL 61/3-1. 
This research has made use of data and/or software provided by the High Energy Astrophysics Science Archive Research Center (HEASARC), which is a service of the Astrophysics Science Division at NASA/GSFC and the High Energy Astrophysics Division of the Smithsonian Astrophysical Observatory. We acknowledge with thanks the variable star observations from the AAVSO International Database contributed by observers worldwide and used in this research.
\end{acknowledgements}

\bibliographystyle{aa}
\bibliography{18516}

\appendix
\section{Optical loading of XMM observations\label{opticalloading}}

Optical loading occurs because the EPIC detectors have a non-zero response to optical photons \citep{lumb00}. Thus, optical photons from a source may generate electrons that would be confused with the charge clouds generated by X-ray photons. For moderately bright optical sources, the net effect is to increase the apparent energy of detected X-ray events. For bright sources, the optical photons that bombard single or adjacent pixels may register as pseudo X-ray events. Three optical blocking filters (thin, medium, and thick) can be employed to reduce optical loading. The limiting magnitude for each blocking filter depends on the spectrum of the source, with redder sources suffering more from optical loading \citep{lumb00}. EPIC-pn slew observations are performed in full frame mode (frame time 0.07 s) with the medium optical blocking filter, and hence, for visually bright and red sources, e.g.~AGB stars, optical loading represents a potential problem. The medium filter is expected to prevent optical contamination from point sources brighter than $V$\,=\,6--9\,mag depending on the color of the source.

To evaluate the likelihood of optical loading for the AGB stars in Table~\ref{lc_vis}, their visual light curves were investigated. Three of the objects found in the XMM slew survey are semi-regular variables (SRb) and one is a Mira (M). All vary up to several magnitudes in the optical during one period. Validated data from the American Association of Variable Star Observers (AAVSO) was used to estimate the visual magnitude of the objects at the time of the X-ray observations. Most of the data available are visual data, i.e. obtained by comparing the brightness of the source to nearby comparison stars by eye. This data has been proven to be accurate within 0.1\,mag \citep{lawsetal90,moonetal08}. To estimate the visual magnitude, $V$, at the time of the X-ray observation, a low order polynomial was fitted (using least-square minimization) to the light curve within $\pm30$\,d of the XMMSS observations. The results are given in Table~\ref{lc_vis} together with the standard deviation, $\Delta V$. Three of the AGB stars are brighter than the optical loading limit of the medium optical blocking filter. Only R~Leo is fainter.

\begin{table}[htdp]
\caption{Visual magnitudes, $V$, with standard deviation errors, $\Delta V$, at the time of the XMMSS X-ray observation, from the AAVSO archive. The last two columns give the results of the optical loading calculations (in photons\,pixel$^{-1}$\,frame$^{-1}$) for an M0 and an M8 source, respectively (see text for details). }
\begin{center}
\begin{tabular}{lccccc}
\hline \hline
Source & JD & $V$ & $\Delta V$ & O.L. for M0 & O.L. for M8 \\
 & & & & [$\gamma$ pix$^{-1}$ fr.$^{-1}$] & [$\gamma$ pix$^{-1}$ fr.$^{-1}$] \\
\hline
W~Cyg & 2454434.2 & \phantom{1}6.1 & 0.3 & \phantom{1}49\phantom{.1} & \phantom{1}335\phantom{.1} \\
R~Dor & 2453831.9 & \phantom{1}6.1 & 0.2 & \phantom{1}49\phantom{.1} & \phantom{1}335\phantom{.1} \\
R~Leo & 2454045.7 & 10.0 & 0.3 & \phantom{10}1.4 & \phantom{100}9.2 \\
L$_{2}$~Pup & 2453069.5 & \phantom{1}6.9 & 0.5 & \phantom{1}23\phantom{.1} & \phantom{1}160\phantom{.1} \\
SS~Vir & 2453532.4 & \phantom{1}8.0 & 0.5 & 490\phantom{.1} & 3360\phantom{.1} \\
\hline
\end{tabular}
\end{center}
\label{lc_vis}
\end{table}

\citet{lumb00} calculates V-band zero magnitude spectral energy distributions (SEDs) for a range of spectral types and convolves the SEDs with a model of the detector/optical blocking filter optical response.  With appropriate scaling by the observed source V-band magnitude, these estimates are used to determine the potential optical contamination. Using the estimated V-band magnitudes, the scaled optical flux (photons\,pixel$^{-1}$\,frame$^{-1}$) for each detected source is estimated and presented in Table~\ref{lc_vis}. \citet{lumb00} did not consider SEDs representative of AGB stars, thus, the range determined from the M0 and M8 spectral type calculations in \citet{lumb00} should be considered as a crude estimate. The observed objects are likely redder and the optical flux is possibly even higher than estimated from these calculations. If each optically-generated electron contributes a charge of 3.6\,eV, then a single-pixel pseudo X-ray event of 0.15\,keV would be generated from a scaled optical flux of 42\,photons\,pixel$^{-1}$\,frame$^{-1}$.    

\begin{figure}[]
\begin{center}
\includegraphics[width=\columnwidth]{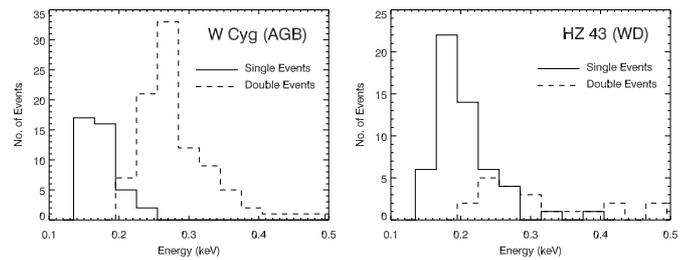}
\caption{These two plots show the energy distributions of the slew survey-detected sources of the AGB star W~Cyg and WD HZ~43.  Single events (pattern number 0, solid lines) are the most reliable events in an XMM X-ray source, while double events (pattern numbers 1-4, dashed lines) are events with charge spread over two pixels. The distribution of single and double events in HZ~43 is the expected distribution for a soft and bright X-ray source, whereas the distributions exhibited by W~Cyg and the other AGB stars detected in the slew survey (L$_2$~Pup, R~Dor, and R~Leo), are indicative of optical loading from these bright red stars.  }
\label{slewspectra}
\end{center}
\end{figure}

Individually, the brightest stars with V-band magnitudes $\sim$\,6 have predicted optical fluxes ($>$\,42\,photons\,pixel$^{-1}$\,frame$^{-1}$) that are consistent with optically-generated pseudo X-ray events. This suspicion is further confirmed by the energy spectrum for single (pattern 0) and double events (patterns 1-4) depicted in Fig.~\ref{slewspectra}. These spectra demonstrate that the single pixel events peak at $\sim$\,0.15\,keV, while the double events peak around 0.3\,keV, or twice that of the single events. This behavior is consistent with both piled-up X-ray photons from a high-rate source of soft X-ray photons, and the optically-generated electron clouds of a visually bright source. However, few X-ray sources are as sharply peaked as that of W~Cyg at the energy resolution of XMM/EPIC, thus the absence of a continuum casts doubt on the veracity of the detected X-ray events. Comparison with the slew detection of the high-rate, soft, X-ray emitting WD, HZ~43 (see Fig.~\ref{slewspectra}, Slew Survey Source: XMMSL1 J131621.7+290553, ObsID: 9083100004, Obs. Date: 2004-06-23), confirms this suspicion. The single events in the authentic X-ray source associated with HZ~43 appear at higher energies, whereas the pseudo events vanish. We therefore conclude that the XMM slew sources associated with our sample objects are due entirely to optical loading by the visually bright red AGB stars.

\subsection{Origin of the X-ray emission detected from SS Vir}
\label{ss_vir}
The detection of SS Vir is highly suspect because it falls on the edge of the FOV, in a region that is normally discarded from science-quality data products, but which may be exposed. The background-subtracted count rate of SS Vir is $4.8\pm1.3$\,counts\,ks$^{-1}$ in the 0.2-2\,keV energy range. The soft spectrum of the source is similar to that seen from the slew sources, thus, as described above, the possibility of optical loading of the SS Vir observation must be investigated. The visual magnitude determined from the AAVSO observations of SS Vir at the time of the XMM observation (see Table~\ref{lc_vis}) suggests SS Vir was clearly bright enough to cause optical loading in the thin optical blocking filter observation. The energy distributions of the high-quality and bad events do not display the high-energy tails seen in the optically loaded slew sources (e.g. Fig.~\ref{slewspectra}); however, the detector optical loading response for bright, red optical sources in this region of the detector is unknown.  

\section{Hydrogen column densities estimated from 21cm-observations}
\label{nh}

\begin{figure}[]
\begin{center}
\includegraphics[width=7cm]{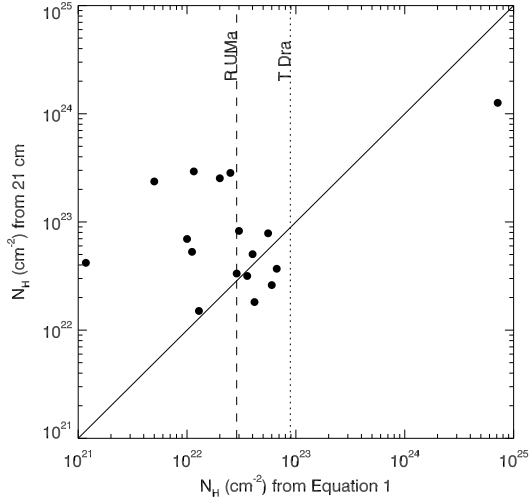}
\caption{Hydrogen column densities estimated from 21cm-observations \citep{geralebe06,mattetal11} compared to estimates from the mass-loss rate.}
\label{hr}
\end{center}
\end{figure}

We have compared the results of the equation given by \citet{kastsoke04a} to estimates from measurements of the 21cm-emission of atomic hydrogen around AGB stars \citep{geralebe06,mattetal11} by calculating the hydrogen column densities from the estimated mass and extent of the emission, assuming an $r^{-2}$-density distribution (Fig.~\ref{hr}). Some sources will be missing at the low $N_{\rm{H}}$ end of the plot because in low-temperature, high-mass-loss rate AGB stars most of the gas will be in molecular form. Both estimates have substantial uncertainties, and although there is a large scatter around the one-to-one-correlation, there is no indication that the equation systematically overestimates the hydrogen column density as suggested by \citet{kastsoke04a}. 

\end{document}